# On the location of the excess wing relative to the α-loss peak in the susceptibility spectra from simulations with the swap Monte Carlo algorithm

K. L. Ngai

*CNR-IPCF, Largo B. Pontecorvo 3, I-56127, Pisa, Italy*

**Abstract**

An advance was made by Guiselin et al. [arXiv:2103.01569 (2021)] in molecular dynamics simulations of the equilibrium dynamics of supercooled liquids near the experimental glass transition by utilizing the giant equilibration speedup provided by the swap Monte Carlo algorithm. The found emergence of a power law in relaxation spectra at lower temperatures on the high frequency flank of the α-loss peak in analogy to the excess wings observed experimentally in molecular glass-formers. Their remarkable finding leads to the question of where the excess wing is located relative to the α-loss peak in the susceptibility spectrum. I provide an answer by identifying the excess wing as the unresolved Johari-Goldstein (JG) β-relaxation and using the reciprocal of its relaxation time $\tau_{JG}$ to assess the location of the excess wing. The Coupling Model (CM) has a history of being successful in determining approximately the values of $\tau_{JG}(T)$ in molecular liquids whether the JG β-relaxation is resolved or not (i.e., excess wing). It is applied to the simulation data and the results of $1/\tau_{JG}(T)$ successfully account for the locations of the excess wings in the simulation spectra at different temperatures. The time evolution of the dynamics of the distribution of processes composing the JG β-relaxation suggested by the CM based on experimental data in molecular glass-formers are in agreement with that given by Guiselin et al. from simulations of a size-polydisperse mixture of N soft spheres interacting with a repulsive power law pair potential. The agreement brings their results closer to experimental data of real molecular glass-formers.

## I. Introduction

An advance in molecular dynamics simulations of the equilibrium dynamics of supercooled liquids was recently reported by Guiselin et al. [1]. The immense equilibration speedup by the swap Monte Carlo algorithm was combined with extensive multi-CPU molecular dynamics simulations [2,3]. The simulations were able to explore the first ten decades of the equilibrium dynamics of supercooled liquids in an effort to approach the glass transition seen by experiments. A size-polydisperse mixture of $N$ soft spheres [1] of mass $m$ interacting with a repulsive power law pair potential was studied in 2 and 3 dimensions. The self-intermediate scattering functions, $F_s(t) = \langle N^{-1} \sum_i \cos[\boldsymbol{q} \cdot \delta \boldsymbol{r}_i(t)] \rangle$, were obtained in 3 dimensions over a range of temperatures, where $\delta \boldsymbol{r}_i(t)$ is the displacement of particle $i$ over time $t$ and $\boldsymbol{q}$ is the wave-vector. The angular brackets indicate the ensemble average over independent runs and an angular average over wave-vectors with $q$=6.9 at the first peak of the structure factor. The structural α-relaxation time $\tau_\alpha$ was defined by $F_s(\tau_\alpha) = 1/e$.

The $F_s(t)$ exhibits a fast initial decay near $t \approx \tau_o$, at all temperatures $T$, which comes from thermal vibrations. At longer times, $F_s(t)$ decays to zero slowly and is well-described by a stretched exponential function,

$$\varphi(t) = \exp[-(t/\hat{\tau}_\alpha)]^\beta \tag{1}$$

with an almost temperature independent stretching exponent $\beta \approx 0.56$ from $T$=0.095 down to 0.0793. The time temperature superposition property was used in conjunction with the stretched exponential function with constant $\beta \approx 0.56$ to estimate $\tau_\alpha$ for $0.07 \leq T \leq 0.0793$, where the final decay of $F_s(t)$ is observable. Finally Guiselin et al. [1] used an Arrhenius law to extrapolate $\tau_\alpha$ to lower temperatures down to the experimental glass temperature, $T_g \approx 0.056$ [2].

They computed the dynamic susceptibility $\chi''(\omega)$ from $F_s(t)$ after it was transformed to a distribution of relaxation times for all temperatures in the range $0.095 \geq T \geq 0.059$ where $F_s(t)$ are simulated. The $\chi''(\omega)$ spectra obtained are reproduced from Fig.1(c) of Guiselin et al. [1] and shown as Fig.1 in the present paper. At all temperatures the spectra display a peak at high frequency $\omega_o \approx 1/\tau_o$, corresponding to the short-time decay of $F_s(t)$. However the low-frequency α-loss peak located at $\omega_\alpha \approx 1/\tau_\alpha$ is captured by the simulation frequency window only at higher temperatures within the range $0.095 \geq T \geq 0.08$. At temperatures below $T$=0.08 where the α-peak is not directly measured, Guiselin et al. extrapolated its frequency dispersion $\chi''_\alpha(\omega)$ by inserting the result calculated from a stretched exponential fit of $F_s(t)$ using with the same exponent β = 0.56 determined at higher temperatures, and with $\tau_\alpha$ given by the Arrhenius extrapolation. The dashed lines in Fig.1 represent the estimated α-loss $\chi''_\alpha(\omega)$ at lower temperatures. It is well known at high frequencies that the frequency dependence of $\chi''_\alpha(\omega)$ derived from the stretched exponential function is proportional to the power law of $\omega^{-\beta}$, which is $\omega^{-0.56}$ and the slope of the dashed lines at high frequencies in Fig.1 in the present case. By contrast, the $\chi''(\omega)$ spectra from the simulations at temperatures below 0.07 show a power law $\omega^{-\sigma}$ (full line in Fig.1) with the exponent σ ≈ 0.38 smaller than β=0.56. Thus the simulated $\chi''(\omega)$ is in excess of the α-loss peak $\chi''_\alpha(\omega)$ at high frequencies analogous to the excess wing found in the experimental dielectric loss spectra of some molecular glass-formers such as glycerol [4-8], propylene carbonate [4,5], cresol phthalein-dimethylether [9], a metallic glass [10], and many more.

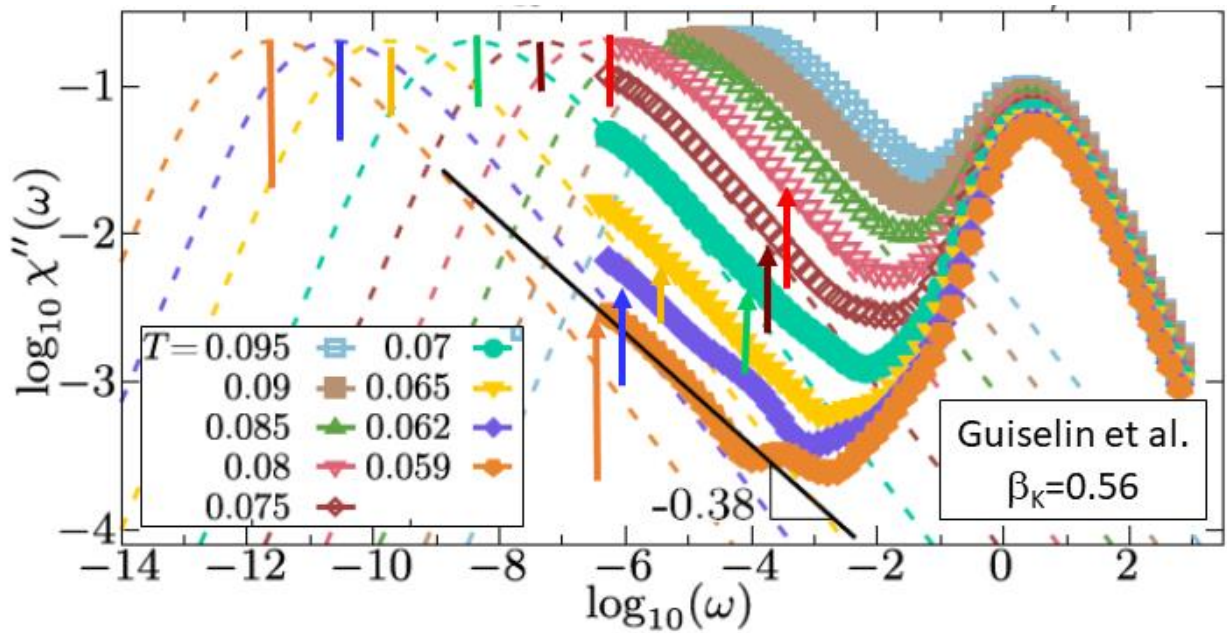

**Figure 1.** Relaxation spectra were obtained by Guiselin et al. [1] by simulations at the temperatures indicated. The dashed lines represent the α-loss peaks corresponding to the Fourier transform of the stretched exponential function with $\beta$=0.56. At lower temperatures close to $T_g$, the spectra show on the high frequency side the presence of a power law proportional to $\omega^{-\sigma}$ with $\sigma \approx 0.38$ (full line), equivalent to the excess wings observed experimentally in real molecular glass-formers. The arrow with the same color as the data indicate the location of the primitive relaxation frequencies $\omega_p(T)$ = $1/\tau_p(T)$ at the corresponding temperature calculated by the CM Eq.(3) (see text).

For comparison I replot in Fig.2 the dielectric loss $\varepsilon''(\nu)$ data of glycerol from Schneider et al. [5] covering an immense frequency range of nearly 18 decades. Shown are the fits of the α-loss peak $\varepsilon''_\alpha(\omega)$ by the Fourier transforms of the stretch exponential relaxation function with constant stretch exponent $\beta$=0.71. The excess wing is revealed by the excess of $\varepsilon''(\nu)$ data over $\varepsilon''_\alpha(\omega)$ contributed by the α-relaxation. The striking similarity of the simulation results in Fig.1 to the experimental data justifies the claim of an advance in simulations of dynamics of glass-formers by Guiselin et al.[1] since the excess wings are directly observed and the microscopic origin can be assessed. In fact they have shown that the combination of spatially heterogeneous dynamics and kinetic facilitation provides a microscopic explanation for the emergence of excess wings in deeply supercooled liquids. Motivated by these findings, a minimal empirical model was constructed by Scalliet et al. [11] to describe this physics and introduce dynamic facilitation in the trap model.

After finding the presence of the excess wing by either simulations or experiment, naturally the relevant question to ask is the location of the excess wing in the susceptibility spectrum relative to the α-loss peak frequency $\omega_\alpha \approx 1/\tau_\alpha$ at different temperatures. This is the question I attempt to answer in this paper.

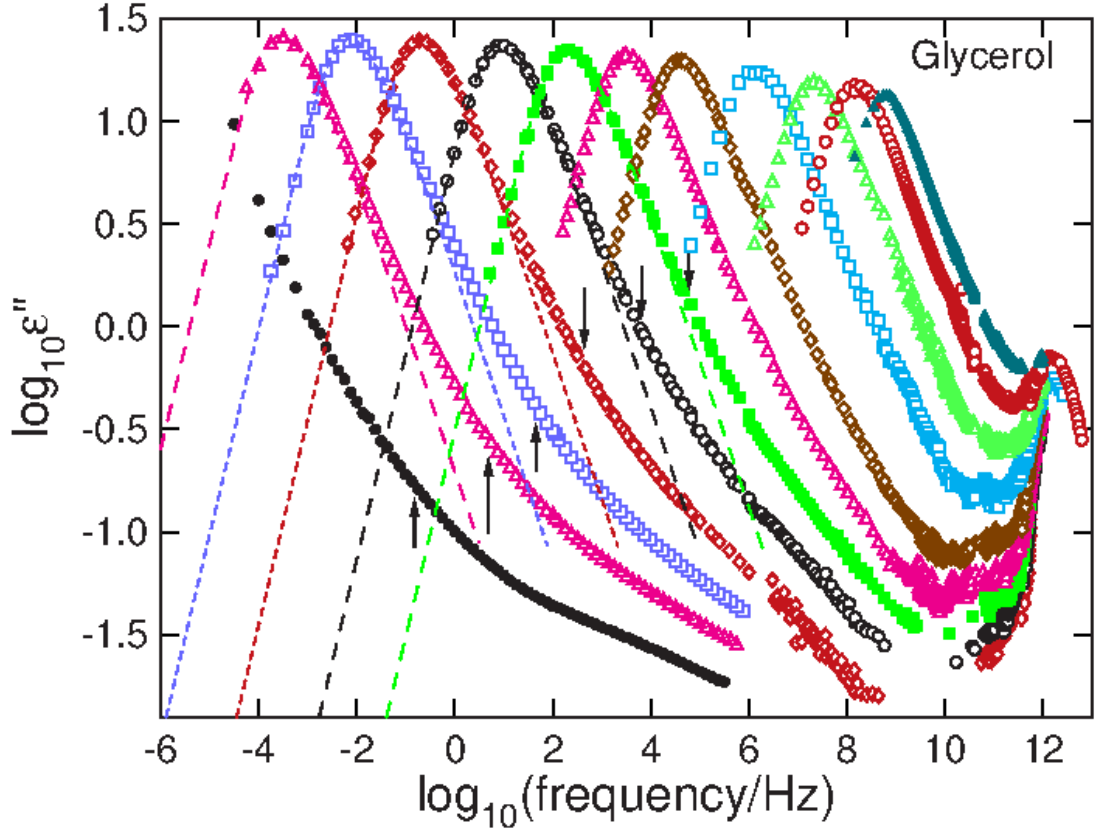

**Figure 2.** Dielectric loss spectra of glycerol at temperatures of 204, 213, 223, 234, 253, 273, 295, and 323 K (from left to right). Data from Schneider et al. [4] are replotted. The dashed lines are the fits of the stretched exponential α-loss peaks by the Fourier transforms of the stretched exponential functions with $\beta$=0.71. The arrows indicate the primitive relaxation frequencies $\nu_p(T)$ via $\tau_p(T)$ which is calculated by the CM Eq.(3).

## II. Results from the Coupling Model

As far as I know, the Coupling Model (CM) is the only theoretical framework having given answer to this question and successfully in doing so [12-14]. The excess wing is an unresolved secondary relaxation of a special kind having strong connection to the α-relaxation in various properties as found by experiments [14-18]. Such secondary relaxation is called the Johari-Goldstein (JG) β-relaxation with relaxation time $\tau_{JG}$ in honor of Johari and Goldstein [19-21] in pioneering studies of secondary relaxations in glass-formers. An example of its strong connection to the α-relaxation is the invariance of the ratio $\tau_\alpha(T,P)/\tau_{JG}(T,P)$ to variations of *T* and pressure *P* while maintaining $\tau_\alpha(T,P)$ constant in van der Waals molecular liquids [14-18]. This property applies to the excess wing in relation to the α-relaxation because the excess wing is just an unresolved JG β-relaxation [13]. The JG β-relaxation observed in simulations and experiments is found to be comprised of processes with different length-scales, cooperativity, and dynamic heterogeneity by experiments [22-24] and by simulations [25-33]. These properties are summarily discussed in Ref.[34] with respect to the viewpoint of CM on the JG β-relaxation. In the CM there is the primitive relaxation which is simplest relaxation involving a single molecule and is a precursor of the distribution of processes in the JG β-relaxation involving participation of increasing number of molecules as time increases. Thus the primitive relaxation is a part of the distribution of processes composing the JG β-relaxation, and hence the primitive relaxation time $\tau_p(T,P)$ is approximately equal to some characteristic JG β-relaxation time $\tau_{JG}(T,P)$ determined from the experiment by some procedure from the distribution, *i.e.*,

$$\tau_{JG}(T,P) \approx \tau_p(T,P). \tag{2}$$

The CM has the equation relating $\tau_\alpha(T,P)$ to $\tau_p(T,P)$,

$$\tau_\alpha(T,P) = [t_c^{\beta-1}\tau_p(T,P)]^{1/\beta}, \tag{3}$$

where $t_c$ is a constant equal to about 1-2 ps for molecular glass-formers, and $(1-n)\equiv\beta$ is the exponent of the stretch exponential function in Eq.(1). For readers not familiar with the long history of the CM since 1979, it is important to note that both the KWW form of the correlation function in Eq.(1) and also the key CM Eq.(3) were derived theoretically from several versions of the CM [11-13]. Also the value of $t_c \approx$1-2 ps for molecular glass-formers and polymers is not ad hoc, and instead was determined directly in neutron scattering experiments [35-38]. The primitive $\tau_p(T,P)$ can be obtained via Eq.(2) from $\tau_\alpha(T,P)$ and stretch exponent *β*, and in turn approximately $\tau_{JG}(T,P)$ from Eq.(2). The final result is $\tau_{JG}(T,P)$ can be calculated from $\tau_\alpha(T,P)$ and *β*, and it is approximately given by

$$\tau_{JG}(T,P) \approx [\tau_\alpha(T,P)]^\beta (t_c)^{1-\beta} = \tau_p(T,P) \tag{4}$$

This approximate or order of magnitude agreement between the calculated $\tau_p(T,P)$ and the experimental $\tau_{JG}(T,P)$ has been verified in many different glass-formers where the JG process is resolved [14-18,22]. In glass-formers where the JG process is unresolved, it appears in the spectra as excess wing such as glycerol in Fig.2. Proofs of the excess wing is indeed the unresolved JG processes come from of long-term aging experiments whereby the excess wing is changed to become a shoulder [5,14], and from mixing with another glass-former with higher $T_g$ resulting in changing the excess wing to a resolved JG β-relaxation [17]. From the glycerol dielectric data in Fig.2, the primitive $\tau_p(T)$ were calculated from $\tau_\alpha(T)$ and stretch exponent *β*=0.71. As indicated by the arrows, the corresponding primitive frequencies $\nu_p(T) =1/2\pi\tau_p(T)$ and hence by Eq.(4) also the JG β-relaxation frequency $\nu_\beta(T)$ fall within the excess wings at several temperatures.

In the same way I perform the CM calculation of the primitive relaxation time for the spectra of the simulations by Guiselin et al.[1] shown in Fig.1. In order to carry this task, the simulation time which was scaled by $\tau_o$ has to be converted to real time. The conversion is based on $10^8$ of the scaled simulation time or $1/\omega$ in Fig.1 corresponds to 10 ms of real time [1]. By this rule, the reciprocals of the simulation peak frequencies $1/\omega_\alpha \approx \tau_\alpha$ of the α-relaxation at a number of temperatures from 0.08 down to 0.059 are converted to real times. The $t_c$ in the CM equation (3) and (4) for the simulated system cannot be determined from the self-intermediate scattering functions $F_s(t)$ of the simulations because $F_s(t)$ exhibits a fast initial decay near $t \approx \tau_o$, at all temperatures $T$, which comes from thermal vibrations. Notwithstanding, the value of $\tau_o \approx 10^{-10}$ s from the simulation is taken as $t_c$ for the system and used in Eq.(4) together with the stretching exponent $\beta \approx 0.56$ to calculate $\tau_p(T)$ in real time. The reciprocals of the calculated values of $\tau_p(T)$ in real time were converted back to primitive frequency $\omega_p(T)$ of simulation, and the values for several temperatures are indicated by the arrows. The $\omega_p(T)$ results in Fig.1 are similar to $\nu_p(T)$ of glycerol in Fig.2. The two primitive relaxation frequencies are located within the principal frequency domain of the excess wing representing the unresolved JG β-relaxation.

It is worthwhile to point out that the separation of the excess wing from the α-loss peak in the simulation spectra (Fig.1) is few decades more than that in the dielectric spectra of glycerol (Fig.2). The separation is defined by $\Delta = (\log\omega_{on} - \log\omega_\alpha)$, where $\omega_{on}$ is the frequency at the onset of deviation of the data from the stretch exponential fit. For an objective comparison it has to be made at the same peak α-loss peak $\omega_\alpha$. At $T$=0.065, $\log\omega_\alpha$=-9.8, which correspond to the log of the real angular frequency ω (in units of s$^{-1}$) equal to 0.2 or $\log(\nu/Hz)$=-0.4. The glycerol spectra at 253 K having $\log(\nu/Hz)$=-0.7 is close enough make the comparison. The value of $\Delta\approx2.7$ for simulation is larger than $\Delta$=0.13 for glycerol. The disparity in $\Delta$ is due to the smaller stretch exponent β equal to 0.56 from simulations than 0.71 for glycerol.

The nature of the JG β-relaxation was elucidated before by experiments [22-24] and simulations [25-33]. The relaxation starts with the single molecule independently and locally, which is the primitive relaxation of the CM [34]. As time increases, more and more molecules relax jointly in heterogeneous domains of increasing in size, and contribute a distribution of relaxation times of the JG. This evolution of dynamics continues until the length scales of these processes approach the cooperative length-scale of the α-relaxation. After that, the α-relaxation starts with time correlation function well described by the stretched exponential function [34]. This description of evolution of dynamics of the processes composing the JG β-relaxation has support from the snapshots in real time of motion of all colloidal particles by confocal microscopy (see Fig.S1 in Supplementary Information) or simulations of Li ions in Li metasilicate glasses (see Fig. S2 in SI), and the metallic glasses [30-33]. More support experiments came from deuteron NMR [23], dielectric hole burning [24], and nuclear resonance synchrotron X-ray scattering [38-42], which were reviewed and discussed in connection with the CM in two recent papers [34,43].

The interpretation of the excess wing as the unresolved JG β-relaxation in terms of the CM has strong support from dielectric studies of the change of the α-relaxation by confining glycerol [44] and prilocaine [45] in nanometer spaces. Extreme nano-confinement drastically reduces the length-scale and cooperativity of the α-relaxation and convert it to the primitive relaxation/JG β-relaxation. The coupling parameter $n\equiv(1-\beta)$ of the CM becomes zero or $\beta$ becomes 1, and it follows from Eqs.(3) and (4) that $\tau_\alpha(T)$ is reduced to $\tau_p(T) \approx \tau_{JG}(T)$. As shown in Refs.[46,47] the CM prediction is indeed verified by the experimental data [44,45]. It is worthwhile to mention in passing the paper by Pabst et al. [48] challenging the CM interpretation of the excess wing and relation to the narrow α-loss peak observed by dielectric spectroscopy in polar liquids including glycerol (see Fig.2). Instead Pabst et al. suggested the dielectric spectrum is the sum of a strong Debye-like contribution from cross-correlation of dipoles and a broad but weak contribution from the self-correlation contribution. This suggestion has been contradicted by experimental data in several papers [49-52], and hence it is invalid.

Since the excess wing is just an unresolved JG β-relaxation, the description of the dynamics of the JG β-relaxation by the CM [14-18] is applicable to the excess wing, and can be compared with the results of the simulations in 2 dimensions by Guiselin et al. [1]. They showed the 2d snapshots of motions of particles as time increases at a temperature $T_{2d} = 0.09$, corresponding to $\tau_\alpha \approx 10$ ms in real time. The description is best by citing in their own words: "… for $t \ll \tau_\alpha$, relaxation starts at a sparse population of localized regions which emerge independently throughout the sample over broadly distributed times" and "as time increases, newly relaxed regions continue to appear, but a second mechanism becomes apparent in Fig. 2 as regions that have relaxed in one frame typically appear larger in the next. The further growth of relaxed regions in Fig. 2 is the signature of dynamic facilitation." Effectively the nature of the dynamic processes in the excess wing found by Guiselin et al. is the same as for resolved JG β-relaxation found by experiments and simulations in colloidal particles [14,22] and glass-formers [14-18]. In the previous works [34,53,54] I warned against the use of the practice of fitting experimental data when the JG β-relaxation is unresolved and appears as the excess wing or shoulder. The practice is to fit the excess wing or shoulder by an empirical function such as the Cole-Cole, adding it to the α-loss peak represented by the Fourier transform of the stretch exponential function, and the sum is used to fit the experimental data. The fault of the practice is that the JG β-relaxation and the α-relaxation are not independent and unrelated processes, and their contributions to susceptibility are not additive. The same point also was made by Guiselin et al. by saying that one should not interpret the excess wing as an additive "β-process", because the small value of the exponent σ would lead to a secondary process much broader than the main α-peak, which seems unphysical.

In glass-formers having resolved JG β-relaxation the remarkable property found is the invariance of the ratio $\tau_\alpha(T,P)/\tau_{JG}(T,P)$ to variations of $T$ and $P$ while $\tau_\alpha(T,P)$ is kept constant. This property is in accord with the CM equation (3) because it turns out that the stretch exponent $\beta(T,P)$ is invariant under the same condition [55]. In glass-formers where the JG β-relaxation is not resolved and appears as excess wing, it was found that essentially the combination of the α-loss peak and the excess wing remain unchanged in shape to variations of $T$ and $P$ while $\tau_\alpha(T,P)$ is kept constant (see Ref.17 and references therein). This property of excess wing would be interesting for further study by the molecular dynamics simulations using the swap Monte Carlo algorithm to investigate in the future.

## Supplementary Information

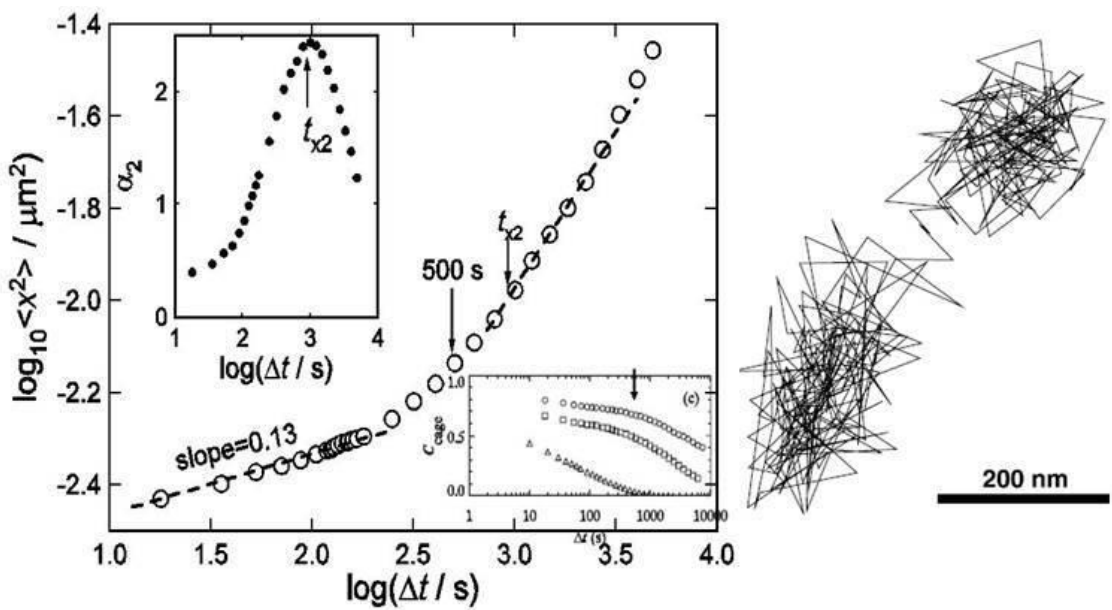

**Figure S1. (Left)** Mean square displacement $\langle\Delta x^2(\Delta t)\rangle$ for volume fractions $\phi$=0.56 from Weeks et al. [S1]. One vertical arrow indicate $\Delta t$ =500 s, the time when a typical particle shifts position and leaves the cage determined by confocal microscopy experiment. The other vertical arrow indicates the time $t_{x2}$ when the non-Gaussian parameter $\alpha_2(\Delta t)$ assumes its maximum as shown in the inset on the upper-left corner. The dashed line has slope 0.13 indicates the nearly constant loss (NCL) regime. The inset on the lower-right corner is a plot of the cage correlation function $C_{cage}(\Delta t)$ against $\Delta t$ for three systems with $\phi$=0.56, 0.52, and 0.46 (from top to bottom), and the vertical arrow indicates $\Delta t$ =500 s. **(Right)** A 2D representation of a typical trajectory in 3D for 100 minutes for $\phi$=0.56 from Weeks et al.[S1] to illustrate that particles spent most of their time confined in cages formed by their neighbors and moved significant distances only during quick

rare cage rearrangements. The particle shown took ~500 s to shift position. Reproduced from Refs.[S1, S2] by permission.

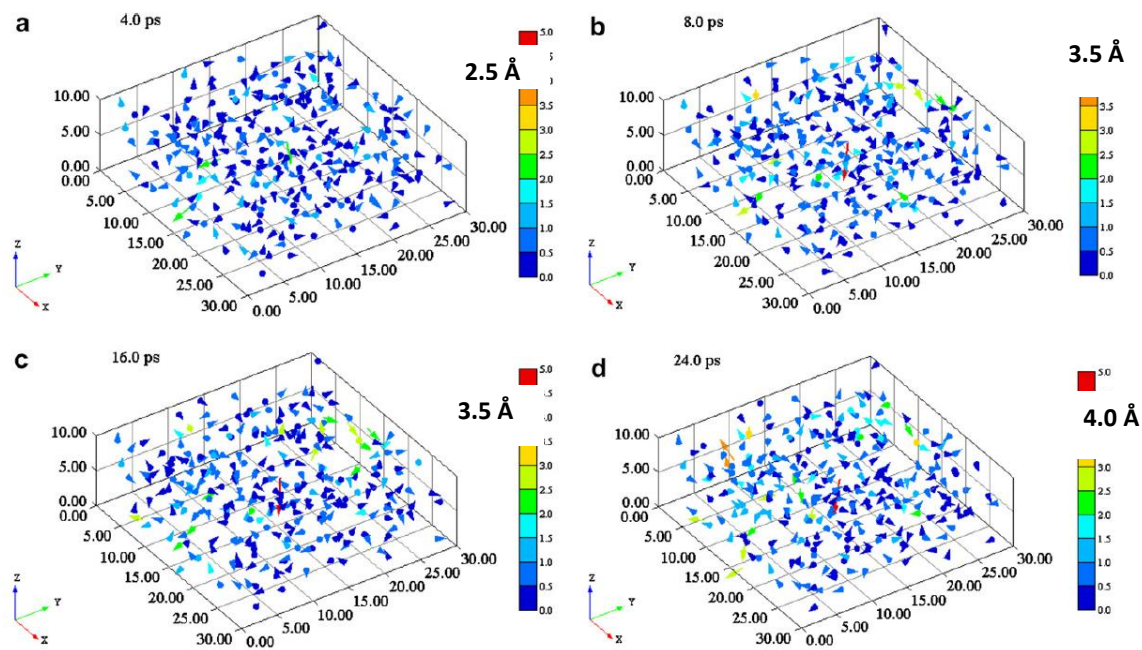

4.5 Å

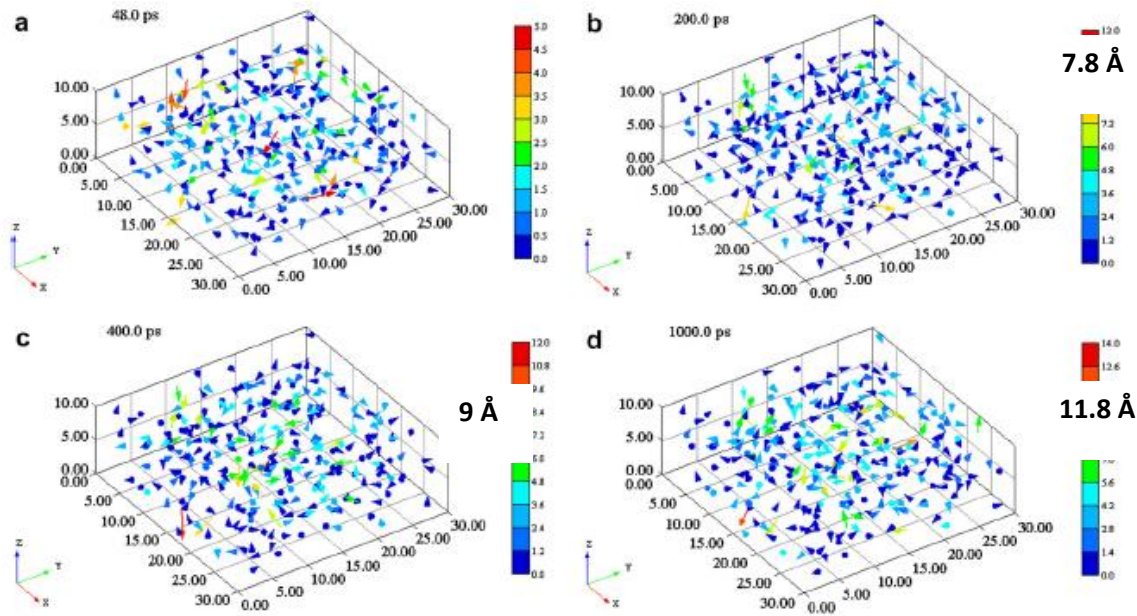

**Figure S2**. **(Upper four)** Motion of Li ions in Li$_2$SiO$_3$ at 700 K at four different times, (a) 4 ps, (b) 8 ps, (c) 16 ps, (d) 24 ps. The positions of the Li ions at any of the indicated chosen times are represented by the vectors from the positions at an initial time in three dimensions for a part of the basic cell of the simulation. The values of axes are in Å. The colors are used to indicate the lengths of the vectors (the values shown in the legend are also in Å). Note that the code of the color scales for 24 ps is different from those at shorter times. In each case, the vertical arrow help to indicate approximately the rare maximum displacement of the Li ions, and the value is recorded on top. It is 2.5, 3.5, 3.5, and 4.0 Å for 4, 8, 16, and 24 ps respectively. But note that most of the displacements have magnitudes smaller than the maximum value. **(Lower four)** Motion of Li ions in Li$_2$SiO$_3$ at 700 K at four times, (a) 48 ps, (b) 200 ps, (c) 400 ps, (d) 1000 ps. The positions of the Li ions at any of the indicated chosen times are represented by the vectors from the positions at an initial time in three dimensions for a part of the basic cell of the simulation. The values of axes are in Å. The colors are used to indicate the lengths of the vectors (the values shown in the legend are also in Å). Note that the code of the color scales for 48 ps and 200 ps, and for 400 ps and 1000 ps are different. In each case, the vertical arrow help to indicate approximately the maximum displacement of the Li ions, and the value is recorded on top. It is 4.5, 7.8, 9.0, and 11.8 Å for 48, 200, 400, and 1000 ps respectively. But note that most of the displacements have magnitudes smaller than the maximum value. Reproduced from Ref.[S3] with permission.

## References

13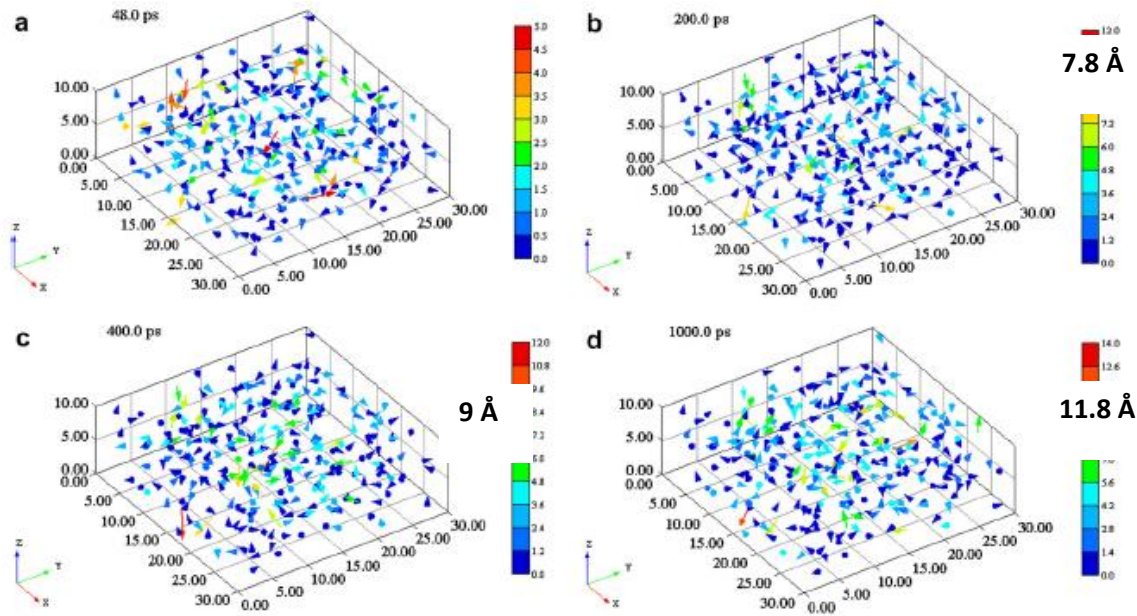

**Figure S2**. **(Upper four)** Motion of Li ions in Li$_2$SiO$_3$ at 700 K at four different times, (a) 4 ps, (b) 8 ps, (c) 16 ps, (d) 24 ps. The positions of the Li ions at any of the indicated chosen times are represented by the vectors from the positions at an initial time in three dimensions for a part of the basic cell of the simulation. The values of axes are in Å. The colors are used to indicate the lengths of the vectors (the values shown in the legend are also in Å). Note that the code of the color scales for 24 ps is different from those at shorter times. In each case, the vertical arrow help to indicate approximately the rare maximum displacement of the Li ions, and the value is recorded on top. It is 2.5, 3.5, 3.5, and 4.0 Å for 4, 8, 16, and 24 ps respectively. But note that most of the displacements have magnitudes smaller than the maximum value. **(Lower four)** Motion of Li ions in Li$_2$SiO$_3$ at 700 K at four times, (a) 48 ps, (b) 200 ps, (c) 400 ps, (d) 1000 ps. The positions of the Li ions at any of the indicated chosen times are represented by the vectors from the positions at an initial time in three dimensions for a part of the basic cell of the simulation. The values of axes are in Å. The colors are used to indicate the lengths of the vectors (the values shown in the legend are also in Å). Note that the code of the color scales for 48 ps and 200 ps, and for 400 ps and 1000 ps are different. In each case, the vertical arrow help to indicate approximately the maximum displacement of the Li ions, and the value is recorded on top. It is 4.5, 7.8, 9.0, and 11.8 Å for 48, 200, 400, and 1000 ps respectively. But note that most of the displacements have magnitudes smaller than the maximum value. Reproduced from Ref.[S3] with permission.

## References

[S1] E. R. Weeks, J. C. Crocker, A. C. Levitt, A. Schofield, D. A. Weitz, Science **287**, 627 (2000).

[S2] E. R. Weeks, D. A. Weitz, Phys. Rev. Lett. **89**, 095704 (2002).